\documentclass[12pt]{article}

\usepackage{amsmath,amssymb}
\usepackage{slashed}
\usepackage{xcolor} %Define colors
\usepackage{graphicx}
\usepackage{url}
\usepackage{cite}

\def\lsim{\mathrel{\rlap{\lower4pt\hbox{\hskip1pt$\sim$}}
  \raise1pt\hbox{$<$}}}
\def\gsim{\mathrel{\rlap{\lower4pt\hbox{\hskip1pt$\sim$}}
  \raise1pt\hbox{$>$}}}

\newcommand{\beq}{\begin{equation}}
\newcommand{\eeq}{\end{equation}}
\newcommand{\bea}{\begin{eqnarray}}
\newcommand{\eea}{\end{eqnarray}}
\newcommand{\nn}{\nonumber \\ }

\newcommand{\hc}{\mathrm{h.c.}}
\newcommand{\eps}{\epsilon}

\newcommand{\cO}{{\cal O}}
\newcommand{\cL}{{\cal L}}
\newcommand{\cM}{{\cal M}}

\graphicspath{{./Figures/}}

\newcommand{\eref}[1]{Eq.~\eqref{eq:#1}} 
\newcommand{\aref}[1]{Appendix~\ref{app:#1}}
\newcommand{\sref}[1]{Section~\ref{sec:#1}}
\newcommand{\tref}[1]{Table~\ref{tab:#1}}

\begin{document}

\vspace*{-2cm}
\begin{flushright}
 LPT Orsay 14-77 \\
\vspace*{2mm}
\today
\end{flushright}

\begin{center}
\vspace*{15mm}

\vspace{1cm}
{\large \bf 
Model-independent precision constraints on dimension-6 operators
} \\
\vspace{1.4cm}

{Adam Falkowski$\,^{a}$ and Francesco~Riva$\,^{b}$}

 \vspace*{.5cm} 
$^a$Laboratoire de Physique Th\'{e}orique, Bat.~210, Universit\'{e} Paris-Sud, 91405 Orsay, France\\
$^b$ Institut de Th\'eorie des Ph\'enom\`enes Physiques, EPFL, 1015 Lausanne, Switzerland 

\vspace*{.2cm} 

\end{center}

\vspace*{10mm} 
\begin{abstract}\noindent\normalsize

We discuss electroweak precision constraints on dimension-6 operators in the effective theory beyond the standard model.  
We identify the combinations of these operators that are constrained by the pole observables (the W and Z masses and on-shell decays) and by the W boson pair production. 
To this end, we define a set of effective couplings of W and Z bosons to fermions and to itself, which capture the effects of new physics corrections. 
This formalism clarifies which operators are constrained by which observable, independently of the adopted basis of operators.     
We obtain numerical constraints on the coefficients of dimension-6 operator in a form that can be easily adapted to any particular basis of operators, or any particular model with new heavy particles.  

\end{abstract}

\vspace*{3mm}

\newpage

%%%%%%%%%%%%%%%%%%%%%%%%%%%%%%%%%%
\section{Introduction} 
\label{sec:intro}

All existing data indicate that, at the weak scale, fundamental interactions respect the local $SU(3) \times SU(2) \times U(1)$ symmetry of the standard model (SM).
The discovery  of a 125 GeV boson at the Large Hadron Collider (LHC)~\cite{Aad:2012tfa,Chatrchyan:2012ufa} and measurements of its production and decay rates vindicate the Brout-Englert-Higgs mechanism, where linearly realized $SU(2) \times U(1)$ symmetry is spontaneously broken to $U(1)$ via a vacuum expectation value (VEV) of the Higgs field. 
It is reasonable to assume that  any new particles, if they exist, are much heavier than the SM particles.
If that is the case, physics at the weak scale can be adequately described by an effective field theory (EFT)  in which the SM Lagrangian is the leading order term and the effects of new physics are encoded in higher-dimensional operators constructed out of the SM fields. 
This way, the EFT framework allows one to parametrize all possible effects of heavy new physics in a systematic expansion in operator dimensions, which is equivalent to an expansion in the mass scale of the new particles.  
Generically, the leading contributions to physical observables are expected from dimension-6 operators. 
 
The first classification of dimension-6 operators was performed in the 80s \cite{Buchmuller:1985jz}, and a complete, non-redundant set was identified  in Ref.~\cite{Grzadkowski:2010es}. 
Much of recent work has been focused on connecting these operators to observables that can be measured in colliders, and to derive experimental  constraints on their coefficients \cite{Pomarol:2013zra,Elias-Miro:2013mua,Corbett:2012ja,Blas:2013ana,Dumont:2013wma,Willenbrock:2014bja,Ellis:2014dva,Gupta:2014rxa,Masso:2014xra,deBlas:2014ula,Ciuchini:2014dea,Ellis:2014huj}. 
 A complete model-independent study  is complicated by the fact that one needs to deal with the large number of free parameters: 76 for flavor-{ universal} dimension-6 operators \cite{Grzadkowski:2010es}, and 2499 for a general flavor structure \cite{Alonso:2013hga}. 
Meanwhile, it is well known that some combinations of these operators are  constrained by electroweak precision observables, in particular by the Z-pole observables at LEP-1, and by the gauge boson pair production at LEP-2, Tevatron, and the LHC.  
It is of utmost importance to identify the existing constraints on dimension-6 operators, and understand their consequences for future new physics searches. 
These constraints have been discussed in the literature  \cite{Han:2004az,Cacciapaglia:2006pk,delAguila:2011zs,Pomarol:2013zra,Elias-Miro:2013mua,Willenbrock:2014bja,Gupta:2014rxa,Ciuchini:2014dea,Ellis:2014huj}, however a general and quantitative  analysis that is easily interpretable in the context of different sets of operators and is therefore easily applicable to different new physics models is still missing.  

In this paper we derive model-independent constraints on dimension-6 operators from precision electroweak observables. 
We assume  that the dimension-6 operators are flavor { universal}, but otherwise we do not introduce any other model-dependent assumptions. 
In particular, all dimension-6 operators can be present simultaneously with arbitrary coefficients. 
Their magnitude is then determined by comparison with experimental data, allowing the validity of the EFT approach to be verified a posteriori.   

Our constrains are based on precision measurements of the Z and W boson masses and on-shell decays (we call it the {\em pole observables}) and of the W-boson pair production. 
These observables have a nice feature that they do not dependent directly on 4-fermion operators, which greatly reduces the number of relevant parameters and  makes the analysis more tractable. 
We derive analytical formulas describing  how these observables depend on the coefficients of dimension-6 operators. 
Rather than choosing a specific basis, we work with a larger, redundant set of operators, such that our results can be easily applied to any of the popular bases. 
We identify the combination of dimension-6 operators that is probed by each observable.
We show that the pole observables constrain 8 combinations of dimension-6 operators, while the W pair production constrains another 3 combinations.
Our results are presented in a basis-independent fashion, and they can be easily adapted to any particular basis. 
Using  these results one can constrain possible effects in indirect new physics searches,  such as the studies of the Higgs boson properties,  that are affected by the same dimension-6 operators. 
%They also constrain specific models with heavy new particles. 

The paper is organized as follows. 
In \sref{EFFL} we introduce the effective Lagrangian relevant for our analysis.
In \sref{EWPT} we discuss the constraints on dimension-6 operators imposed by the pole observables.   
In \sref{TGC} we discuss further constraints on these operators from the W-boson pair production in LEP-2. 
In \aref{bases} we show how to connect our general results to constraints on dimension-6 operators in specific bases of operators.

%%%%%%%%%%%%%%%%%%%%%%%%%%%%%%%%%%
\section{Effective Lagrangian} 
\label{sec:EFFL} 

We consider an effective theory where the SM is extended by dimension-6 operators: 
\beq
\cL_{\rm eff}  = \cL_{\rm SM}  +   \cL_{D=6}.
\eeq 
% where $v = 246.2$~GeV is the Higgs field VEV. 
The SM Lagrangian in our notation takes the form 
\bea 
{\cal L}_{\rm SM} &=& 
  -{1 \over 4 g_s^2} G_{\mu \nu}^a G_{\mu \nu}^a  - {1 \over 4 g_L^2} W_{\mu \nu}^i W_{\mu \nu}^i  -  {1 \over 4 g_Y^2}    B_{\mu \nu} B_{\mu \nu} 
  + D_\mu H^\dagger D_\mu H  + \mu_H^2 H^\dagger H - \lambda (H^\dagger H)^2  
\nn  &+& 
i \sum_{f \in q, \ell} \bar f \bar \sigma_\mu D_\mu f +  i \sum_{f^c \in u^c,d^c,e^c} f^c  \sigma_\mu D_\mu \bar f^c   
 -  \left [ H  q Y_u u^c +   H^\dagger q  Y_d d^c +  H^\dagger \ell Y_\ell e^c    + \hc \right ].
\eea
The gauge couplings of $SU(3) \times SU(2) \times U(1)$ are denoted by $g_s$, $g_L$, $g_Y$, respectively;  we also define the electromagnetic coupling $e = g_L g_Y/\sqrt{g_L^2 + g_Y^2}$, and the Weinberg angle $\sin \theta_W = g_Y/\sqrt{g_L^2 + g_Y^2}$. 
Note that we use the convention where  the gauge kinetic terms are normalized by the corresponding gauge coupling. 
The Higgs field gets the VEV $\langle  H \rangle = (0,v/\sqrt{2})$. 
After electroweak symmetry breaking, the gauge mass eigenstates are defined as 
$W^\pm = (W^1 \mp W^2)/(\sqrt 2 g_L)$, $Z = (W^3 - B)/\sqrt{g_L^2 + g_Y^2}$, $A =  (g_Y^2 W^3 + g_L^2 B)/g_L g_Y \sqrt{g_L^2 + g_Y^2}$.
For the fermions we use the 2-component notation, with all conventions as in Ref.~\cite{Dreiner:2008tw}. 

We are interested in the subset of dimension-6 operators that contribute to electroweak precision observables. To avoid the complication of dealing with a large number of parameters, in this paper we restrict to observables that are {\em not} sensitive to 4-fermion operators\footnote{%
In practice, one 4-fermion operator enters in our analysis via a shift of the input parameters, since it affects the measurements of $G_F$ via the muon decay rate, see below \eref{EWPT_chats}. This can be avoided if one uses $m_Z,m_W$ and $\alpha$ to fix the SM input parameters \cite{Gupta:2014rxa}.}, such as the decay widths of on-shell W and Z bosons and the pair production of W and Z bosons. 
Typically, at this point one makes a choice of a {\em  basis}, that is of a non-redundant set of operators relevant for  studied processes. 
Our goal in this paper is to discuss electroweak precision constraints on dimension-6 operators in a way that can be easily adapted to any of the popular bases.  
For this reason, we will work with a redundant set of operators, and identify the combination of operators that are constrained by precision observables.   
The relevant operators  are given by
\bea
\label{eq:EFFL_ld6pole}
\cL_{D=6}  & \supset & 
 {c_T \over 4 v^2} H^\dagger \overleftrightarrow { D_\mu} H  H^\dagger \overleftrightarrow {D_\mu} H  
 + {c_{WB} \over 4 m_W^2}  B_{\mu\nu}  W_{\mu\nu}^i H^\dagger \sigma^i H  
 + i {c_{HW}  \over m_W^2}  D_\mu H^\dagger\sigma^i D_\nu H W^i_{\mu\nu} 
  + i {c_{HB}   \over m_W^2}  D_\mu H^\dagger D_\nu H B_{\mu\nu}
 \nn  &+ & 
 i {c_W \over 2 m_W^2}  H^\dagger  \sigma^i \overleftrightarrow  {D_\mu} H  D_\nu  W_{\mu \nu}^i
 + i {c_B \over 2 m_W^2}  H^\dagger  \overleftrightarrow {D^\mu} H \partial_\nu  B_{\mu \nu}
 + {c_{2W} \over 16 m_W^2} (D_\rho  W_{\mu \nu}^i)^2 
  + {c_{2B} \over 16 m_W^2} (\partial_\rho  B_{\mu \nu})^2 
\nn  &+ & 
 i {c_{HQ}'  \over v^2} \bar q \sigma^i \bar \sigma_\mu q H^\dagger \sigma^i \overleftrightarrow {D_\mu} H   
+ i{ c_{HQ}  \over v^2}\bar q \bar \sigma_\mu q H^\dagger  \overleftrightarrow {D_\mu} H  
+ i {c_{HU}  \over v^2}  u^c \sigma_\mu \bar u^c  H^\dagger  \overleftrightarrow {D_\mu} H 
+i {c_{HD}  \over v^2}  d^c \sigma_\mu \bar d^c  H^\dagger  \overleftrightarrow {D_\mu} H 
\nn  &+ &  
i {c_{HL}'  \over v^2} \bar \ell \sigma^i \bar \sigma_\mu \ell H^\dagger \sigma^i \overleftrightarrow {D_\mu} H   
+i {c_{HL} \over v^2} \bar \ell \bar \sigma_\mu \ell H^\dagger  \overleftrightarrow {D_\mu} H  
+i {c_{HE}  \over v^2} e^c \sigma_\mu \bar e^c  H^\dagger  \overleftrightarrow {D_\mu} H
 + {c_{3W} \over 6 g_L^2 m_W^2} \epsilon^{ijk}    W_{\mu \nu}^{i} W_{\nu\rho}^{j} W_{\rho \mu}^{k},
 \nn
\eea 
where  $H^\dagger  \overleftrightarrow {D_\mu} H = H^\dagger D_\mu H - D_\mu H^\dagger H$. 
In the following we will often shorthand these operators by $O_X$ defined via ${\cal L}^{D=6} \equiv c_X O_X$.
The operators in \eref{EFFL_ld6pole} form a redundant set because they can be related by  equations of motion: 
\bea
\label{eq:EFFL_eqofmo}
{1 \over g_Y^2} \partial_\nu B_{\mu \nu} &=&
{i \over 2} H^\dagger  \overleftrightarrow {D_\mu} H  +     \sum_f Y_f \bar f \bar \sigma_\mu f -    \sum_{f} Y_{f^c} f^c  \sigma_\mu \bar f^c, 
\nn
{1 \over g_L^2} D_\nu W_{\mu \nu}^i &=& 
{i \over 2} H^\dagger \sigma^i  \overleftrightarrow {D_\mu} H  +    {1 \over 2} \sum_f \bar f \sigma^i \bar \sigma_\mu f. 
\eea
Using these, one finds the operators $O_W$ and $O_B$ are equivalent to a combination of $O_T$ and the vertex operators $O_{HF}$  
Similarly, $O_{2W}$ and $O_{2B}$ can be traded for other operators in \eref{EFFL_ld6pole} and 4-fermion operators.  
%where the operators $O_X$ are defined via ${\cal L}^{D=6} \equiv c_X O_X$. 
Moreover, certain operators in  \eref{EFFL_ld6pole} can be related by integration by parts: 
\beq
\label{eq:EFFL_inbypa}
O_{HB} =  O_B - O_{WB} - O_{BB}, \qquad O_{HW} =  O_W - O_{WB} - O_{WW},
\eeq    
%\bea
 %i D_\mu H^\dagger D_\nu H B_{\mu\nu} &=&  
%{i \over 2} H^\dagger  \overleftrightarrow {D_\mu} H \partial_\nu  B_{\mu \nu} 
 % - {1 \over 4} H^\dagger \sigma^i H W_{\mu \nu}^i B_{\mu\nu} - {1 \over 4} H^\dagger H B_{\mu\nu} B_{\mu \nu}
% \nn 
% i D_\mu H^\dagger\sigma^i D_\nu H W^i_{\mu\nu}  &=& 
% {i \over 2} H^\dagger  \sigma^i \overleftrightarrow {D_\mu} H D_\nu  W_{\mu \nu}^i 
 % - {1 \over 4} H^\dagger \sigma^i H W_{\mu \nu}^i B_{\mu\nu} - {1 \over 4} H^\dagger H W_{\mu\nu}^i W_{\mu \nu}^i
%\eea  
where the operators $O_{BB} =  {1 \over 4 m_W^2} H^\dagger H B_{\mu\nu} B_{\mu \nu}$, $O_{WW} =  {1 \over 4 m_W^2} H^\dagger H W_{\mu\nu}^i W_{\mu \nu}^i$ affect only  Higgs decays but not  electroweak precision observables, therefore they are not included in \eref{EFFL_ld6pole}.
The relations in \eref{EFFL_eqofmo} and \eref{EFFL_inbypa} imply that only linear combinations of these operators affect physical observables. 
A choice of a basis consists in picking a non-redundant subset of these operators, such that any single operator can be constrained by experiment.  
For example, in the so-called {\em Warsaw basis} of Ref.~ \cite{Grzadkowski:2010es} the operators $O_W$,  $O_B$, $O_{2W}$, $O_{2B}$, $O_{HW}$, and $O_{HB}$ are dropped,
while in the {\em SILH basis} \cite{Giudice:2007fh,Contino:2013kra}, the operators $O_{HL}'$, $O_{HL}$ and $O_{WB}$ are dropped.
Specific bases are discussed in more detail in \aref{bases}. 

The operators in \eref{EFFL_ld6pole} contribute to precision observables in a three-fold way.
Firstly, the operators   $O_T$, $O_{WB}$, $O_{W}$,  $O_{B}$, $O_{2W}$, and $O_{2B}$ affect the propagators of electroweak gauge bosons (the so-called {\em oblique} corrections). 
We define these via the 2-point functions of the SM gauge bosons 
$\cM(V_{\mu} \to V_{\nu}) =   \eta_{\mu\nu}  \Pi_{V V}(p^2)  + p_{\mu}p_{\nu}   ( \dots )$, 
and the momentum expansion $ \Pi_{V V}(p^2) = \Pi_{V V}^{(0)} +  \Pi_{V V}^{(2)} p^2  +   \dots$.  
The oblique corrections $\delta \Pi_{VV}$ are deviations of the propagator functions from the canonical form. 
In the mass eigenstate basis the oblique corrections are related to those in the electroweak basis by  
\bea
\delta \Pi_{WW} &=& g_L^2 \delta  \Pi_{W^1 W^1} = g_L^2 \delta  \Pi_{W^2 W^2},
\nn 
\delta \Pi_{ZZ} &=& {1 \over g_L^2 + g_Y^2} \left (g_L^4 \delta \Pi_{W^3 W^3}  -2 g_L^2 g_Y^2 \delta \Pi_{W^3B}  + g_Y^4 \delta \Pi_{BB}  \right),
\nn 
\delta \Pi_{\gamma \gamma} &=& {g_L^2 g_Y^2 \over g_L^2 + g_Y^2} \left ( \delta \Pi_{W^3 W^3}  +2 \delta \Pi_{W^3B}  +  \delta \Pi_{BB}  \right),
\nn
\delta \Pi_{Z\gamma} &=& {g_L g_Y  \over g_L^2 + g_Y^2} \left (g_L^2 \delta \Pi_{W^3 W^3}  +(g_L^2 -  g_Y^2) \delta \Pi_{W^3 B}  - g_Y^2 \delta \Pi_{33}  \right).
\eea 
By electromagnetic gauge invariance,  $\delta \Pi_{B B}^{(0)} =  - \delta \Pi_{W^3 B}^{(0)} = \delta \Pi_{W^3 W^3}^{(0)}$. 
The dimension-6 operators in \eref{EFFL_ld6pole} contribute to the oblique corrections as 
 \bea 
\label{eq:EFFL_dpi}
& \displaystyle \delta \Pi_{W^3 W^3}^{(0)} =  -  {c_T v^2 \over 8},  
\quad 
\delta \Pi_{W^3 B}^{(2)}  = - {c_{WB} +  c_W +  c_B \over g_L^2},
& \nn  &  \displaystyle
\delta \Pi_{W^i W^i}^{(2)} ={2 c_W  \over g_L^2}, \quad  \delta \Pi_{B B}^{(2)}  = {2 c_B  \over g_L^2}, \quad 
\delta \Pi_{W^i W^i}^{(4)} ={c_{2 W}  \over g_L^2 v^2}, \quad  \delta \Pi_{B B}^{(4)}  = {c_{2 B}  \over g_L^2 v^2}. 
\eea
The shift of the diagonal kinetics terms of  by the operators $O_{W}$, $O_{B}$ has no physical consequences but it's important to keep track of,  to properly read off the contributions to gauge boson self-interactions.  

Another effect on precision observables arises due to a shift of the couplings of W and Z bosons to fermions.   
In general, the interactions of electroweak gauge bosons with the SM fermions can be parametrized as  
\bea
\label{eq:EFFL_vertex_par}
\cL_{ffV} &=&  e A_\mu \sum_{f = u,d,e} Q_f \left ( \bar f \bar \sigma_\mu f  +  f^c   \sigma_\mu \bar f^c \right )
\nn 
&+& {g_L \over \sqrt 2} W_\mu^+   \left [ 
(1 + \delta g_{qW,L}) \bar u \bar \sigma_\mu   V_{\rm CKM} d 
+ (1 + \delta g_{\ell W,L}) \bar e \bar \sigma_\mu  \nu \right ] + \hc 
\nn 
&+& \sqrt{g_L^2 + g_Y^2} Z_\mu 
\sum_{f = u,d,e,\nu}   (T^3_f - \sin^2 \theta_W Q_f  + \delta g_{fZ,L}) \bar f \bar \sigma_\mu  f  
\nn 
&+& \sqrt{g_L^2 + g_Y^2} Z_\mu 
\sum_{f = u,d,e}   (- \sin^2 \theta_W Q_f  + \delta g_{fZ,R}) f^c  \sigma_\mu  \bar f^c     
\eea 
The operators in \eref{EFFL_ld6pole}  induce the following vertex corrections\footnote{% 
There is another dimension-6 operator $ic_{HUD} u^c \sigma_\mu \bar d^c   H  {D_\mu} H + \hc$ leading to  the vertex correction  $\delta g_{qW,R} W_\mu^+  u^c  \sigma_\mu  \bar d^c  + \hc$. 
However, this operator does not interfere with the SM and thus contributes to observables only  at the quadratic level, therefore we ignore it here.  
}
\bea & 
\label{eq:EFFL_dg}  \displaystyle
\delta g_{qW,L}  = c_{HQ}' ,  \qquad \delta g_{\ell W,L}  = c_{HL}',  
& \nn &   \displaystyle
\delta g_{uZ,L}  = {c_{HQ}' \over 2} - {c_{HQ} \over 2},
\quad  
\delta g_{dZ,L}  =  - {c_{HQ}' \over 2} -  {c_{HQ} \over 2}, 
\quad  
\delta g_{uZ,R}  =  - {c_{HU} \over 2},
\quad  
\delta g_{dZ,R}  = -  {c_{HD} \over 2},
& \nn &   \displaystyle
\delta g_{\nu Z,L}  = {c_{HL}' \over 2} - {c_{HL} \over 2},
\quad 
\delta g_{e Z,L}  = - {c_{HL}' \over 2} -  {c_{HL} \over 2}, 
\quad  
\delta g_{e Z,R}  =  - {c_{HE} \over 2}. 
\eea 

Finally, dimension-6 operators affect WW and WZ pair production by contributing to anomalous triple gauge couplings (TGCs). 
In the customary parametrization  in Ref.~\cite{Hagiwara:1986vm,Hagiwara:1993ck,De Rujula:1991se}:    
 \bea 
 \label{eq:EFFL_tgcpar}
 \cL_{TGC}  &= &  
i  e (1 -  \delta \Pi_{W^i W^i}^{(2)})  \left [  
  \left (  W_{\mu \nu}^+ W_\mu^-  -  W_{\mu \nu}^- W_\mu^+ \right ) A_\nu  + (1 + \delta \kappa_\gamma)  A_{\mu\nu}\,W_\mu^+W_\nu^- \right ] 
\nn  &  + & i g_L \cos \theta_W (1 -  \delta \Pi_{W^i W^i}^{(2)} )  \left [  
 (1 + \delta g_{1,Z} )  \left ( W_{\mu \nu}^+ W_\mu^-  -  W_{\mu \nu}^- W_\mu^+ \right ) Z_\nu 
 +  \left (1 +\delta \kappa_Z  \right )  \, Z_{\mu\nu}\,W_\mu^+W_\nu^-
  \right ] 
  \nn & + &    
  i e \frac{ \lambda_\gamma}{m_W^2} W_{\mu \nu}^+W_{\nu \rho}^- A_{\rho \mu}  
  +  i g_L \cos \theta_W     \frac{ \lambda_Z}{m_W^2} W_{\mu \nu}^+W_{\nu \rho}^- Z_{\rho \mu}, 
\eea 
where the factor $\delta \Pi_{W^i W^i}^{(2)}$ (which cancels in  physical observables) arises because modifications  of the kinetic term of the $SU(2)$ gauge bosons by gauge symmetry imply the corresponding shift of TGCs. 
The operators in \eref{EFFL_ld6pole} contribute to the anomalous TGCs as\footnote{{ Note that we take the signs of TGCs in \eref{EFFL_tgcpar} opposite to that of Ref.~\cite{Hagiwara:1986vm} because we use a different convention for the covariant derivatives: $D = \partial - igV$.}}
\beq  
\delta  g_{1,Z} =  - {g_L^2 + g_Y^2 \over g_L^2} \left ( c_W + c_{HW} \right ), 
\quad  
\delta \kappa_\gamma =  c_{WB} - c_{HW} - c_{HB},  
\quad 
\lambda_Z =  -   c_{3W}. 
\eeq  
while $\lambda_\gamma = \lambda_Z$, and $\delta \kappa_Z =  \delta g_{1,Z} -  {g_Y^2 \over g_L^2} \delta \kappa_\gamma$. 

In the rest of this paper we discuss the current constraints from pole observables and gauge boson pair production on the  dimension-6 operators in \eref{EFFL_ld6pole}.  

%%%%%%%%%%%%%%%%%%%%%%%%%%%%%%%%%%%%%%%%
\section{Precision constraints on Z and W pole} 
\label{sec:EWPT}

In this section we discuss the constraints on dimension-6 operators from  precision observables that involve a single on-shell Z or W boson. 
We refer to them jointly as the {\em pole observables}. 
In order to confront these observables with the SM predictions, numerical values of the electroweak parameters in the SM have to be determined from some input observables.  
As is the common practice, for the input observables we take the muon decay width $\Gamma(\mu \to e \nu \nu)$ (directly related to the Fermi constant $G_F = 1/\sqrt{2}v^2$),  the low-energy electromagnetic constant $\alpha(q^2 = 0)$,  and the Z boson mass $m_Z$.  With this choice, the electroweak parameters take the values $g_L = 0.657$,  $g_Y =  0.341$, $v = 246.2$~GeV.

 \begin{table}
 \begin{center}
 \begin{tabular}{|c|c|c|c|c|}
\hline
{\color{blue}{Observable}} & {\color{blue}{Experimental value}}   &   {\color{blue}{Ref.}}   &  {\color{blue}{SM prediction}}    &  {\color{blue} Definition}  
  \\  \hline   \hline
$m_Z$ [GeV]  & $91.1875 \pm 0.0021$  &   \cite{ALEPH:2005ab} &   $\times$   &  $\sqrt{{(g_L^2 + g_Y^2) v^2 \over 4}  + \delta \Pi_{ZZ}(m_Z^2)}$
\\ 
\hline 
$\Gamma_{Z}$ [GeV]  & $2.4952 \pm 0.0023$ & \cite{ALEPH:2005ab} & $ 2.4950$    & $\sum_f \Gamma (Z \to f \bar f)$ 
 \\  \hline
$\sigma_{\rm had}$ [nb]  & $41.540\pm 0.037$ &  \cite{ALEPH:2005ab} &  $41.484$ &  ${12 \pi \over m_Z^2} {\Gamma (Z \to e^+ e^-) \Gamma (Z \to q \bar q) \over \Gamma_Z^2}$ 
  \\  \hline 
 $R_{\ell}$  & $20.767\pm 0.025$ & \cite{ALEPH:2005ab} &  $20.743$  &    $ {\sum_{q} \Gamma(Z \to q \bar q) \over  \Gamma(Z \to \ell^+ \ell^-)} $
 \\  \hline
 $A_\ell$ & $0.1499 \pm 0.0018$ &  \cite{Baak:2014ora} &  $0.1472$ &  ${\Gamma(Z \to e_L^+ e_L^-) - \Gamma(Z \to e_R^+ e_R^-) \over \Gamma(Z \to e^+  e^-) }$  \\ \hline
 $A_{\rm FB}^{0,\ell}$ & $0.0171\pm 0.0010$ &  \cite{ALEPH:2005ab} &  $0.01626$   &  ${3 \over 4} A_\ell^2$  
 \\  \hline
$R_b$ & $0.21629\pm0.00066$ & \cite{ALEPH:2005ab} & $0.21578$  &     ${ \Gamma(Z \to d \bar d) \over \sum_q \Gamma(Z \to q \bar q)}$ 
   \\  \hline
$A_b$ & $0.923\pm 0.020$ & \cite{ALEPH:2005ab} & $0.93463$  &
 ${ \Gamma(Z \to d_L \bar d_L)  -   \Gamma(Z \to d_R \bar d_R)  \over  \Gamma(Z \to d \bar d)   }$ 
 \\  \hline
$A_{b}^{\rm FB}$ & $0.0992\pm 0.0016$ & \cite{ALEPH:2005ab}  & $0.1032$  & ${3 \over 4} A_\ell A_b$  \\  \hline
$R_c$ & $0.1721\pm0.0030$  & \cite{ALEPH:2005ab}  & $0.17226$  
& ${ \Gamma(Z \to u \bar u) \over \sum_q \Gamma(Z \to q \bar q)} $ 
\\  \hline
$A_c$ & $0.670 \pm 0.027$ & \cite{ALEPH:2005ab} & $0.668$ 
 &  ${ \Gamma(Z \to u_L \bar u_L)  -   \Gamma(Z \to u_R \bar u_R)  \over  \Gamma(Z \to u \bar u)   }$ 
\\  \hline 
$A_{c}^{\rm FB}$ & $0.0707\pm 0.0035$  &   \cite{ALEPH:2005ab} &  $0.0738$  & ${3 \over 4} A_\ell A_c$ 
\\ \hline \hline 
 $m_{W}$ [GeV]  & $80.385 \pm 0.015$ &  \cite{Group:2012gb}    &  $80.364$   &  $\sqrt{{g_L^2 v^2 \over 4}  + \delta \Pi_{WW}(m_W^2)}$
\\ \hline  
$\Gamma_{W}$ [GeV]  & $ 2.085 \pm 0.042$  & \cite{Beringer:1900zz} &  $2.091$  &  $ \sum_f  \Gamma(W \to f f')$   \\
\hline
${\rm Br} (W \to {\rm had})$ & $ 0.6741 \pm 0.0027$ &  \cite{Schael:2013ita} &  $0.6751$  &  $ {\sum_q \Gamma(W \to q q') \over  \sum_f  \Gamma(W \to f f')}$
 \\ \hline \hline 
\end{tabular}
\end{center}
\caption{
The {\em pole observables} used in this analysis. 
We take into account the experimental correlations between the LEP-1 Z-pole observables and between the heavy flavor observables.  
For the theoretical predictions we use the best fit SM values from GFitter \cite{Baak:2014ora}, except for ${\rm Br} (W \to {\rm had})$ where we take the value quoted in  \cite{Schael:2013ita}. 
There's no SM prediction for $m_Z$ because we use it as an input to determine the SM parameters. 
We do not include $\sin^2 \theta^\ell_{\rm eff} (Q_{\rm FB})$ because of the difficulties to interpret this measurement in the presence of vertex corrections.}
\label{tab:EWPT_pole}
 \end{table}

The LEP, SLC, and Tevatron experiments  precisely measured the mass and the total widths of the Z and W boson.   
Moreover, LEP-1 and SLC  measured  relative rates and asymmetries of Z decays into leptons and hadron. 
In \tref{EWPT_pole} we summarize the pole observables used in this analysis, and provide their expression  in terms of the  Z and W partial decay widths into SM  fermions.  
Assuming flavor blind couplings and no new light particles that W and Z can decay into, there are 9 independent partial widths, all of which can be extracted from the pole observables. 
In particular, decays into left- and right-handed fermions can be experimentally separated thanks to the forward-backward and polarization asymmetry measurements. 

The W and Z partial widths together with the W mass measurement make 10  pole observables (in our formalism, the Z boson mass is used as an input to determine the SM parameters, therefore it does not provide constraints on new physics). 
However, the number of independent constraints is smaller:  
it turns out that the pole observables constrain only 8 combinations of dimension-6 operators. 
Specifically, we will show that all pole observables depend on the coefficients of the operators in \eref{EFFL_ld6pole} only via the combinations of parameter $\hat c$ defined as 
\bea
\label{eq:EWPT_chats}
\hat c_{HL}'   &=& c_{HL}'   + c_{WB}+ c_W + c_B  - {g_L^2 \over  4 g_Y^2} c_T + {1 \over 4} c_{2W} + {g_Y^2 \over 8 g_L^2} c_{2B},
\nn 
\hat c_{HL}   &=& c_{HL}   -  { 1 \over 4} c_T -  {g_Y^4 \over 8 g_L^4} c_{2B},
\nn 
\hat  c_{HE} & = &  c_{HE}  - {1 \over 2} c_T -  {g_Y^4 \over 4 g_L^4} c_{2B}, 
\nn
\hat c_{HQ}' &=& c_{HQ}'  + c_{WB}+ c_W + c_B   - {g_L^2 \over 4 g_Y^2 } c_T + {1 \over 4} c_{2W} + {g_Y^2 \over 8 g_L^2} c_{2B},
\nn  
\hat c_{HQ} &=& c_{HQ}  + {1 \over 12} c_T +  {g_Y^4 \over 24 g_L^4} c_{2B}, 
\nn 
\hat c_{HU} &=&  c_{HU} + {1 \over 3} c_T  +  {g_Y^4 \over 6 g_L^4} c_{2B}, 
\nn 
\hat c_{HD} &=&  c_{HD}  - {1 \over 6} c_T -   {g_Y^4 \over 12 g_L^4} c_{2B}, 
\nn
\hat c_{ll} &=& c_{ll} + {1 \over 2} c_{2W}
\eea 
where $c_{ll}$ is the coefficient of the 4-fermion operator  $O_{ll}= - v^{-2} (\bar e \bar \sigma_\rho \nu_e) (\bar \nu_\mu \bar \sigma_\rho \mu)$ in the effective Lagrangian.
This 4-fermion operator enters indirectly, via the contribution to the muon decay width, which is one of our input observables.
Contributions of all other  4-fermion operators  to the  pole observables are suppressed by $\Gamma_Z/m_Z$ or $\Gamma_W/m_W$,  therefore  they can be neglected at the leading order. 

Let us discuss how the pole observables listed  in \tref{EWPT_pole} depend on the coefficients of dimension-6 operators.  
One kind of observables are the physical masses of the W and Z boson.
In the presence of new physics corrections these are given by $m_W = \sqrt{{g_L^2 v^2 \over 4}  + \delta \Pi_{WW}(m_W^2)}$, $m_Z = \sqrt{{(g_L^2 + g_Y^2) v^2 \over 4}  + \delta \Pi_{ZZ}(m_Z^2)}$. 
The effect of the dimension-6 operators  on the oblique corrections can be read off from \eref{EFFL_dpi}. 
Moreover, one should also take into account that new physics contributing to our input observables effectively shifts the SM electroweak parameters $g_L$, $g_Y$ and $v$: 
\bea 
\label{eq:EWPT_input_shift}
{\delta g_L \over g_L} &=& {1 \over g_L^2 - g_Y^2} \left (   
 2{ \delta  \Pi_{WW}^{(0)} \over v^2}  
 - 2 \cos^2 \theta_W    { \delta  \Pi_{ZZ}(m_Z^2)  \over v^2} 
+  {g_Y^2 \over 2}   \delta  \Pi_{\gamma \gamma}^{(2)}  
- g_L^2 \delta g_{\ell W,L} 
+ { g_L^2  c_{ll} \over 4}
\right ),
\nn
 {\delta g_Y \over g_Y} &=& {1 \over g_L^2 - g_Y^2}  \left (   
- {2 g_Y^2  \over g_L^2 }   {\delta  \Pi_{WW}^{(0)} \over v^2} 
+ 2 \sin^2 \theta_W  { \delta  \Pi_{ZZ}(m_Z^2)  \over v^2} 
-  {g_L^2 \over 2 }   \delta  \Pi_{\gamma \gamma}^{(2)}  
 + g_Y^2 \delta g_{\ell W,L} 
- {g_Y^2  c_{ll} \over 4}
\right ),
\nn
 {\delta v \over v}  &=&  -{ 2 \delta  \Pi_{WW}^{(0)} \over g_L^2 v^2}   
+ \delta g_{\ell W,L}
- {c_{ll} \over 4}, 
\eea 
Using  \eref{EFFL_dpi} and  \eref{EWPT_input_shift} one finds $\delta m_Z = 0$, while the W mass is shifted by  
\beq
\delta m_W = - {m_W g_Y^2 \over g_L^2 - g_Y^2} \left ( \hat c'_{HL}   - { \hat c_{ll}  \over 4} \right ). 
\eeq  
The remaining pole observables are related to the W and Z partial decays widths.
These are  given by  $\Gamma(Z \to f \bar f) = {N_f m_Z \over 24 \pi} g_{fZ,\rm eff}^2$,  $\Gamma(W \to f f') = N_f {\hat m_W \over 48 \pi}  g_{fW, \rm eff} ^2$,
where $N_f$ is the number of colors of the fermion $f$. 
The effective couplings are defined as  (see e.g. \cite{Wells:2005vk})
\bea
\label{eq:EWPT_geff} 
g_{fZ ; \rm eff} &=&  { \sqrt{g_L^2 + g_Y^2} \over \sqrt{ 1 - \delta \Pi_{ZZ}'(m_Z^2)}} \left [T^3_f -  Q_f \sin^2 \theta_W \left (  1  - {g_L \over g_Y} {\delta \Pi_{\gamma Z}(m_Z^2) \over m_Z^2} \right )   + \delta g_{fZ} \right ], 
\nn 
 g_{fW; \rm eff}  &=& {g_L  ( 1 + \delta g_{fW} ) \over \sqrt{ 1 - \delta \Pi_{WW}'(m_W^2)}}.   
\eea 
such that they capture new physics effects on the vertices and propagators of electroweak gauge bosons. 
At the linear level, new physics shifts the partial widths as   $\delta \Gamma(Z \to f \bar f)  = {N_f m_Z \over 12 \pi } g_{fZ} \delta g_{fZ;\rm eff}$, 
$\delta \Gamma(W \to f f') = { N_f m_W \over 24 \pi} g_L  \delta g_{fW;\rm eff}$,  
where $g_{fZ}   = \sqrt{g_L^2 + g_Y^2} (T^3_f - \sin^2 \theta_W Q_f )$ is the SM Z coupling to $f$, 
and $\delta g_{fZ;\rm eff} = g_{fZ;\rm eff} - g_{fZ}$. 
Using \eref{EFFL_dpi} and \eref{EFFL_dg} we can trade the oblique and vertex correction in \eref{EWPT_geff} for the coefficients of dimension-6 operators. 
For the $Z$-boson couplings we find 
\bea
\label{eq:EWPT_dg}
\delta g_{\nu Z,L; \rm eff}  &=&  - {\sqrt{g_L^2 + g_Y^2} \over 2}
\left (\hat c_{HL}  -  { \hat c_{ll} \over 4}  \right ) ,
\nn 
\delta g_{e Z,L ;\rm eff}  &=&  {\sqrt{g_L^2 + g_Y^2}  \over g_L^2 - g_Y^2}  \left (
 g_Y^2 \hat c'_{HL}  -  {(g_L^2 - g_Y^2) \hat c_{HL} \over 2}  -  {(g_Y^2 + g_L^2)  \hat c_{ll} \over 8} 
 \right ) ,
\nn
\delta g_{e Z,R; \rm eff}  &=&  {\sqrt{g_L^2 + g_Y^2}  \over g_L^2 - g_Y^2}  \left (
g_Y^2 \hat c'_{HL}   -  {(g_L^2 - g_Y^2) \hat c_{HE} \over 2}  - {g_Y^2  \hat c_{ll} \over 4} 
  \right ) ,
\nn 
\delta g_{u Z,L; \rm eff}  &=&  {\sqrt{g_L^2 + g_Y^2}  \over g_L^2 - g_Y^2}  \left (
 -  {(3 g_L^2 + g_Y^2) \hat c'_{HL} \over 6} 
+  {(g_L^2 - g_Y^2) (\hat c'_{HQ} - \hat c_{HQ}) \over 2}   +  {(3 g_L^2 + g_Y^2)   \hat c_{ll} \over 24} 
  \right ), 
\nn 
\delta g_{u Z,R; \rm eff}  &=&  {\sqrt{g_L^2 + g_Y^2}  \over g_L^2 - g_Y^2}  \left (
  - {2 g_Y^2  \hat c'_{HL} \over 3}  -  {(g_L^2 - g_Y^2) \hat c_{HU} \over 2}    + {g_Y^2  \hat c_{ll}  \over 6} 
  \right ) ,
\nn 
\delta g_{d Z,L; \rm eff}  &=&  {\sqrt{g_L^2 + g_Y^2}  \over g_L^2 - g_Y^2}  \left (
  {(3 g_L^2 - g_Y^2) \hat c'_{HL} \over 6} 
-  {(g_L^2 - g_Y^2) (\hat c'_{HQ} + \hat c_{HQ}) \over 2}  -  {(3 g_L^2 - g_Y^2)   \hat c_{ll} \over 24} 
  \right ) ,
\nn 
\delta g_{d Z,R; \rm eff}  &=&  {\sqrt{g_L^2 + g_Y^2}  \over g_L^2 - g_Y^2}  \left (
{g_Y^2 \hat c'_{HL} \over 3}  -  {(g_L^2 - g_Y^2) \hat c_{HD} \over 2} -   {g_Y^2  \hat c_{ll}  \over 12}  
  \right ) ,
\eea
and 
\bea
\delta g_{ \ell W,L, \rm eff}  &=& - {g_L \over g_L^2 - g_Y^2}
 \left ( g_Y^2 \hat c'_{HL} -   {g_L^2  \hat c_{ll} \over 4}  \right ), 
 \nn 
  \delta g_{ q W,L, \rm eff}  &=&  - {g_L \over g_L^2 - g_Y^2}
 \left ( g_L^2  \hat c'_{HL}   -  (g_L^2 -   g_Y^2) \hat c'_{HQ}   -  {g_L^2  \hat c_{ll}  \over 4}  \right ),
\eea
for the $W$-boson couplings to fermions. 
This explicitly demonstrates that precisely 8 combinations of the dimension-6 operators in \eref{EFFL_ld6pole} and the 4-fermion operator $O_{ll}$ can be constrained by the pole observables. 
Clearly, only combinations of fermionic and purely bosonic operators are constrained, but not the two separately. 
The technical reason is  that operators containing fermions  can be traded for purely bosonic operators using the equations of motion (\ref{eq:EFFL_eqofmo}). 
In particular, two combinations of vertex operators can be traded for the operators $O_W$ and $O_B$.
The latter do not contribute to fermion couplings to W and Z, and they contribute to oblique corrections in the same way as $O_{WB}$. 
Therefore,  these two combinations of vertex operators cannot be probed by the  pole observables~\cite{Grojean:2006nn,Gupta:2014rxa}.

We now move to deriving constraints on the dimension-6 Lagrangian from a global fit to the pole observables.  
We construct a $\chi^2$ function from the observables listed in \tref{EWPT_pole}.
Using \eref{EWPT_dg},  we compute corrections to the observables in terms of the relevant combinations of the parameters in the dimension-6 Lagrangian. 
We take into account the correlations between the observables given in \cite{ALEPH:2005ab}. 
Then we minimize the  $\chi^2$ function with respect to $\hat c_{HF}$ and $c_{ll}$. 
With this procedure, we obtain the  following constraints: 
\beq
\label{eq:EWPT_c_pole} 
\left ( \begin{array}{c} 
\hat c_{HL}'   \\
\hat c_{HL}    \\ 
\hat c_{HE}    \\ 
\hat c_{HQ}'   \\ 
\hat c_{HQ}    \\ 
\hat c_{HU}          \\ 
\hat c_{HD}             \\   
\hat c_{ll}                           
\end{array} \right )  = 
\left ( \begin{array}{c} 
-1.9 \pm    1.1 \\ 
1.1  \pm   0.7  \\ 
0.1   \pm 0.6    \\ 
-4.7   \pm  1.9   \\ 
 0.2     \pm    2.0  \\ 
7.0          \pm 6.9   \\ 
-31.3          \pm  10.3\\  
- 4.7     \pm  3.5         
\end{array} \right )   \cdot 10^{-3}, \
\rho = \left ( \begin{array}{cccccccc} 
1 &  -0.49 &  0.31 &  0.17 & -0.05 &   -0.03 &  -0.04    & 0.89\\ 
\cdot   & 1  & 0.42 &     0.08 &  0.00  &  0.06  &  -0.12   & -0.76\\ 
\cdot   & \cdot &  1 &  -0.04  &  -0.09 &    0.09  &  -0.32  & 0.03  \\ 
\cdot   & \cdot & \cdot & 1 &  -0.39  &  -0.73 &  0.59   & 0.01 \\ 
\cdot   & \cdot & \cdot & \cdot & 1 &   0.43 &  0.22  & -0.04 \\ 
\cdot   & \cdot & \cdot &\cdot &  \cdot &  1 & -0.15  &-0.01  \\ 
\cdot   & \cdot & \cdot & \cdot & \cdot & \cdot & 1  & -0.06 \\
\cdot   & \cdot & \cdot & \cdot & \cdot & \cdot & \cdot  & 1 
\end{array} \right ) 
\eeq 
Using these central values $\hat c^0$, the 1-sigma errors  $\delta \hat c$ and the correlation matrix $\rho$ one can reconstruct the $\chi^2$ function for the pole observables as a function of the coefficients of dimension-6 operators: 
$\chi^2_{\rm pole} = \sum_{ij} (\hat c_i - \hat c_i^0)\sigma^{-2}_{ij}  (\hat c_j - \hat c_j^0)$, where $\sigma^{-2}_{ij} =  [\delta \hat c_i \rho_{ij} \delta \hat c_j]^{-1}$. 
If only a subset of the operators is generated in a particular model, the $\chi^2$ function can be minimized with a smaller number of parameters, and new limits valid in this restricted case can be obtained. 
Thus, \eref{EWPT_c_pole} and  \eref{EWPT_chats} allow one to quickly derive the constraints from the pole observables on any model with new heavy particles. 

Clearly, the combinations of dimension-6 parameters defined in \eref{EWPT_chats} are tightly constrained by the pole observables. 
In particular, the combinations involving leptonic vertex corrections are constrained at the level $\cO(10^{-3})$, while those involving right-handed quark  are constrained at the level of $\cO(10^{-2}-10^{-3})$.\footnote{The preference for a non-zero value of $\hat c_{HD}$ is driven by the well-known $2.5\sigma$ anomaly in the  forward-backward asymmetry of $b$-quark production at LEP-1.}
In any basis, the coefficients of dimension-6 operators must either be very small, or tightly correlated so as to satisfy the constraints $\hat c_{HF} \approx 0$.  
Larger new physics corrections are allowed only on the hyper-surface in the operator space where  these constraints are satisfied. 
We refer to this hyper-surface   as  the {\em flat directions of the pole observables}. 

\eref{EWPT_dg} shows the possibility to parametrize the effects of the dimension-6 Lagrangian, using only the modifications of the $Z$-couplings to fermions. Indeed, it is possible, using field redefinitions proportional to the equations of motions and by taking appropriate linear combinations of the dimension-6 operators, to obtain a non-redundant operator basis in which all propagator corrections vanish, $\delta \Pi_{VV}=0$, and there are only  vertex corrections $\delta g_{f Z}$ \cite{Gupta:2014rxa} (modifications to the $W$ couplings are related to the $Z$ couplings by an accidental custodial symmetry at the level of the dimension-6 Lagrangian, $\delta g_{l L,W}=\delta g_{\nu L,Z}- \delta g_{eL,Z}$, $\delta g_{q L,Z}=\delta g_{u L,Z}- \delta g_{dL,Z}$). Such parametrization is particularly useful to compare with experiments, and we will further discuss it in Appendix \ref{app:bases_BSMP}.
 
In the next section we discuss model-independent  constraints  on these flat directions from  vector boson pair production at LEP-2 and the LHC

%%%%%%%%%%%%%%%%%%%%%%%%%%%%%%%%%%%%%%%%%%%
\section{Constraints from electroweak gauge boson pair production at LEP-2} 
\label{sec:TGC} 

The $e^+ e^- \to W^+W^-$ process was studied at LEP-2 at several center-of-mass energies. 
The total cross sections and differential distributions in the $W$ scattering angle are reported in Ref.~\cite{Schael:2013ita}.
In principle, from these measurements one can extract different tensor structure of gauge bosons self-couplings and separate the t- and s-channel photon and Z contributions,  thanks to their different angular and energy dependence.   

% in analogy with W and Z effective couplings on the pole defined in \eref{EWPT_geff}, 
Our first step is to understand which combinations of dimension-6 operators are constrained by the WW production.
To this end we define a set of effective couplings that fully describe the   $e^+ e^- \to W^+W^-$ process  in the presence of new physics.     
One simplifying assumption we introduce at this point is  that  there are only up to  $p^2$ corrections to the gauge boson propagators.\footnote{ 
This is true for most of the operators in \eref{EFFL_ld6pole},  except for $O_{2W}$, $O_{2B}$. 
Therefore, in  the rest of this section we will assume that, using equations of motion,  these two have been traded for other operators  in \eref{EFFL_ld6pole} and 4-fermion operators.  
Dropping these operators greatly  simplifies the discussion of oblique corrections to the WW production, and avoids dealing with the complicated tensor structure of gauge boson self-interactions introduced by $O_{2W}$.}
This implies $\delta \Pi_{VV}(m_V^2) = \delta \Pi_{VV}^{(0)} +  m_W^2 \delta \Pi_{VV}^{(2)}$, and $\delta \Pi_{VV}'(m_V^2) = \delta \Pi_{VV}^{(2)}$.   

The $e^+ e^- \to W^+W^-$ amplitude can be split into t- and s-channel contributions: $\cM = \cM_t + \sum_{V=\gamma,Z} \cM_{s}^V$.  
The first piece is the t-channel neutrino exchange amplitude:  
\beq
\cM_t = - {g_{\ell W ,L; \rm eff }^2 \over 2  t}   \bar \eps_\mu (p_{W^-}) \bar \eps_\nu(p_{W^+}) 
\bar y(p_{e^+})  \bar \sigma_\nu \sigma\cdot(p_{e^-}- p_{W^-}) \bar \sigma_\mu  x(p_{e^-}), 
\eeq 
where $t = (p_{e^-}- p_{W^-})^2$, $\eps$'s are the polarization vectors of $W^\pm$, and $x$, $y$ are the spinor wave-functions of $e^\pm$ (see Ref.~\cite{Dreiner:2008tw}).  
The effective W coupling to leptons $g_{\ell W ,L; \rm eff }$ is defined in \eref{EWPT_geff}, and it includes the effects of vertex corrections and W wave-function renormalization due to oblique corrections.      
It is the same coupling that determines the W decay width into leptons, therefore this part of the amplitude depends on the same combination of dimension-6 operators as the pole observables.  

The remaining part of the amplitude describes the s-channel photon and $Z$ exchange: 
\beq
\cM_{s}^V =  - {1 \over s - m_V^2} 
\left [ g_{e V,L; \rm eff} \bar y(p_{e^+}) \bar \sigma_\rho x(p_{e^-})    +  g_{e V,R;\rm eff}  x (p_{e^+})  \sigma_\rho \bar y(p_{e^-}) \right ] 
   \bar \eps_{\mu}(p_{W^-}) \bar  \eps_{\nu}(p_{W^+}) F_{\mu \nu \rho}^V , 
\eeq
where $s =  (p_{e^-} + p_{e^+})^2$. 
For the photon diagram,  the effective coupling is $g_{e \gamma; \rm eff} = e_{\rm eff}  \equiv {e \over \sqrt {1 - \delta \Pi_{\gamma \gamma}^{(2)}}}$ for both left- and right-handed fermions. 
One finds $\delta e_{\rm eff} = 0$: the photon couplings to matter are not affected by dimension-6 operators. 
For the Z boson diagram,  the effective couplings $g_{e Z; \rm eff}$, defined in \eref{EWPT_geff}, 
are again the same as the ones that determine the Z-boson decay widths into left- and right-handed leptons. 
Qualitatively new effects of dimension-6 operators  enter via the gauge boson vertex function: 
\bea 
F_{\mu \nu \rho}^V &= &   g_{1,V; \rm eff}  \left [ 
\eta_{\rho \mu} p_{W^-}^\nu  - \eta_{\rho \nu} p_{W^+}^\mu + \eta_{\mu \nu} (p_{W^+} - p_{W^-})_\rho \right ]  
\nn &+&
 \kappa_{V;\rm eff} \left [\eta_{\rho \mu} (p_{W^+} + p_{W^-})_{\nu}  - \eta_{\rho \nu} (p_{W^+} + p_{W^-})_{\mu} \right] 
\nn &+&
 { g_{VWW} \lambda_{V}  \over m_W^2} \left [ 
 \eta _{ \rho \mu} \left (  p_{W^+} (p_{W^+} + p_{W^-})  p_{W^-}^\nu - p_{W^+}    p_{W^-} (p_{W^+} + p_{W^-})_{\nu} \right)
 \right . \nn & + & \left .
\eta _{ \rho \nu} \left (  p_{W^+}  p_{W^-} (p_{W^+} + p_{W^-})_{\mu} - p_{W^-} (p_{W^+} + p_{W^-}) p_{W^+}^\mu  \right) \right ]. 
\eea
where $g_{\gamma WW} = e$, $g_{ZWW} = g_L \cos \theta_W$.  
The {\em effective TGCs} in the first two lines are defined  as 
\bea
\label{eq:TGC_tgceff}
g_{1,\gamma; \rm eff} &= & e_{\rm eff}, \quad 
 \kappa_{\gamma; \rm eff} =e_{\rm eff}  \left [ 1  + \delta \kappa_\gamma \right ],  \quad 
\nn 
g_{1,Z; \rm eff}  &=  & {g_L \cos \theta_W \over \sqrt{ 1 - \delta \Pi_{ZZ}^{(2)} }}  
\left [ 1  + {g_L g_Y \over g_L^2 + g_Y^2} \delta \Pi_{\gamma Z}^{(2)}   \right ] \left [1 + \delta g_{1,Z} \right ],
\nn
 \kappa_{Z; \rm eff} &=&  { g_L \cos \theta_W   \over \sqrt{ 1 - \delta \Pi_{ZZ}^{(2)} }}  
\left [ 1+  {g_L g_Y \over g_L^2 + g_Y^2} \ \delta \Pi_{\gamma Z}^{(2)} \right ] \left [1 + \delta \kappa_{Z} \right ]. 
\eea 
These effective TGCs can be directly related to differential distributions that are experimentally observable (unlike the TGCs in the Lagrangian of \eref{EFFL_tgcpar} \cite{Trott:2014dma}). 
By calculating how they depend on the coefficients of  dimension-6 operators we can find out which combinations of dimension-6 operators are probed by the WW production process. 
In the presence of dimension-6 operators the effective TGCs are shifted away from the SM value by 
\bea
\label{eq:TGC_dtgc}
{\delta g_{1,Z; \rm eff} \over  g_L \cos \theta_W }
 \equiv \delta \hat g_{1,Z} & = &  \left (g_L^2 + g_Y^2 \right) 
 \left [ {c_{WB} + c_B - c_{HW} \over g_L^2} - {c_T \over 4 g_Y^2}  - {\hat c_{HL}  - c_{ll}/4  \over g_L^2 - g_Y^2}  \right ],
 \nn 
 {\delta \kappa_{\gamma; \rm eff} \over  e } \equiv \delta \hat \kappa_\gamma &= &  c_{WB} - c_{HW} - c_{HB},
 \nn 
 \lambda_Z &=& -c_{3W}, 
\eea
\beq 
\label{eq:TGC_dtgc2}
\delta g_{1,\gamma; \rm eff} = 0, \quad  {\delta \kappa_{Z,\rm eff} \over  g_L \cos \theta_W  }   =  \delta \hat g_{1,Z} -  {g_Y^2 \over g_L^2} \delta \hat \kappa_\gamma, \quad 
   \lambda_{\gamma} =  \lambda_Z.
\eeq 
We can see that the WW production is sensitive to 3 new combinations of dimension-6 operators appearing in $\delta \hat g_{1,Z}$,  $\delta \hat \kappa_\gamma$,  and  
$\lambda_Z$  in \eref{TGC_dtgc}. 
At the dimension-6 level, all other new physics corrections can be  expressed either by these three combinations ($\delta \kappa_{Z,\rm eff}$  and  $\lambda_{\gamma}$ in  \eref{TGC_dtgc2}) or by the combinations that enter in the pole observables ($\delta g_{\ell W ,L; \rm eff }$,  $\delta g_{\ell Z ,L; \rm eff }$, and  $\delta g_{\ell Z ,R; \rm eff }$). 
For vanishing oblique and vertex corrections, the shifts of our  effective TGCs in \eref{TGC_dtgc}  reduce to the usual anomalous TGCs defined by \eref{EFFL_tgcpar}, which  are commonly used in the literature to parameterize the vector boson pair production.
However, our formulation is more general and is also valid in the presence of oblique and vertex corrections. 
It can be used with any basis of dimension-6 operators,  also when some anomalous TGCs do not appear in that basis.
For example, in the Warsaw basis of Ref.~\cite{Grzadkowski:2010es}, at first sight the anomalous TGC $\delta g_{1,Z}$ does not seem to receive any direct contribution from new physics, as none of the operators in this basis contains the structure appearing in Eq.~(\ref{eq:EFFL_tgcpar}).
Instead, a combination of  vertex and oblique corrections has exactly the same effect as $\delta g_{1,Z}$, which is captured by our formalism. 
The analogous formalism applies to the WW production at the LHC, with $\delta g_{\ell W;\rm eff}$,  $\delta g_{\ell Z; \rm eff}$ replaced by the effective  W and Z couplings to quarks.

Thus,  the WW production provides qualitatively new information about higher-dimensional operators in the effective Lagrangian that cannot be extracted from the pole observables alone.  
We now discuss, at the quantitative level, the constraints on dimension-6 operators from the $e^- e^+ \to W^+W^-$ production data collected by the LEP-2 experiment. 
We take into account the total and differential  production cross section at different center-of-mass energies, as reported in Ref.~\cite{Schael:2013ita}. 
In principle, the $e^- e^+ \to W^+W^-$ process probes 6 combinations of dimension-6 operators: 
$\delta \hat g_{1,Z}$,  $\delta \hat \kappa_\gamma$,  and   $\lambda_Z$  in \eref{TGC_dtgc}, as well as    $\delta g_{\ell W;\rm eff}$,  $\delta g_{\ell Z,L; \rm eff}$, and $\delta g_{\ell Z,R; \rm eff}$ in \eref{EWPT_dg}.  
Using the $e^- e^+ \to W^+W^-$ data we could constrain these 6 combinations, and then combine these constrains with the ones obtained from  the pole observables. 
In practice, however, a simpler procedure is adequate. 
The constraints from the pole observables  imply $\delta g_{W,\ell;\rm eff} \approx  \delta g_{Z,\ell;\rm eff}  \lesssim {\cal O}(10^{-3})$, while the accuracy of the LEP-2 WW measurements is worse, roughly ${\cal O}(10^{-2})$.  
Therefore, for the sake of fitting the WW data, it is a very good approximation to assume $\hat c_{HL}' = \hat c_{HL} = \hat c_{HE} = \hat c_{ll} = 0$, which implies $\delta g_{W,\ell;\rm eff} = \delta g_{Z,\ell;\rm eff} =0$.  
Then one can focus only on the deformations of the SM along the EFT directions defined by  $\delta \hat g_{1,Z}$, $\delta \hat \kappa_{\gamma}$, and $\lambda_Z$, which are unconstrained by the pole observables.
This simplified procedure is equivalent to fitting the three anomalous TGCs $\delta g_{1,Z}$, $\delta \kappa_{\gamma}$, and $\lambda_Z$ in \eref{EFFL_tgcpar}, assuming vanishing oblique and vertex correction.  
From that 3-dimensional fit, using \eref{TGC_dtgc}, one can read off constraint on the coefficients of dimension-6 operators in any basis.
Results of the fits in some particular bases are given in \aref{bases}; below, we only give the results in the language of the anomalous TGCs. 
Our formalism of effective couplings that are directly connected to observable quantities addresses the concerns raised in Ref.~\cite{Trott:2014dma}. 
As a cross-check, we also performed a complete fit where the full non-redundant set of operators contributing to the pole observables and WW production was allowed to vary freely. 
Numerically, the results of that fit are very close to the results of the simplified 3-dimensional TGC fit quoted below, thus validating our procedure. 

To perform the fit, we computed the relevant WW cross sections analytically  as a function of  $\delta g_{1,Z}$, $\delta \kappa_{\gamma}$, and $\lambda_Z$. 
We also included the constraints on the closely related process of single on-shell W boson production in association with a forward electron and a neutrino \cite{Schael:2013ita}. 
In this case, the corrections due to anomalous TGCs are determined numerically using {\tt aMC@NLO}  \cite{Alwall:2014hca}.    
For the SM predictions we take the numbers quoted in \cite{Schael:2013ita}. 
At the {\em linear level} in dimension-6 operators, we find the constraints  
\beq 
\label{eq:TGC_limits_lep2} 
\delta  g_{1,Z} = -0.83 \pm 0.34, \  
\delta \kappa_{\gamma} = 0.14 \pm 0.05, \ 
\lambda_{Z} = 0.86 \pm 0.38,  \quad  
\rho = \left ( \begin{array}{ccc} 
1 & -0.71  & -0.997 \\
\cdot & 1 & 0.69  \\
\cdot & \cdot & 1   
\end{array} \right ) . 
\eeq
The constraints are weaker than expected given the LEP-2 precision, with ${\cal O}(1)$ TGCs allowed by \eref{TGC_limits_lep2}. 
This is  related to the  approximately blind direction of the LEP-2 WW data along $\lambda_Z \approx -\delta g_{1.Z}$  that was pointed out in Ref.~\cite{Brooijmans:2014eja}.\footnote{Indeed, along the direction $\delta\kappa^\pm\equiv (\lambda_Z\pm\delta g_{1.Z})/\sqrt{2}$, one finds that $\delta\kappa^+=0.005\pm0.055$ while $\delta\kappa^-=1.11\pm0.57$.} Notice that this blind direction appears to be a complete accident that occurs for the energy range and the observables explored by LEP-2. 
In particular, for $s \gg (200\textrm{GeV})^2$, the linear level differential cross-section is sensitive separately to $\lambda_Z$  and $\delta g_{1.Z}$. 
Furthermore, the blind direction appears only after summing over the polarizations of $e^\pm$ and $W^\pm$, whereas including polarization information would remove the blind direction.
Single $W$ production data (omitted in  \cite{Brooijmans:2014eja}) do not remove this blind direction because they constrain mostly $\delta \kappa_\gamma$. 
Including in the cross-section the quadratic terms we obtain
 \footnote{%
Note that, in the fits performed so far by experimental collaborations, the quadratic terms are always included.}
\beq
\label{eq:TGC_limits_lep2q}
\delta  g_{1,Z} =  -0.05^{+0.05}_{-0.07}, \quad \delta  \kappa_{\gamma} = 0.05^{+0.04}_{-0.04}, \quad \lambda_Z = 0.00^{+ 0.07}_{-0.07}. 
\eeq
The errors are much smaller than for the linear fit in  \eref{TGC_limits_lep2}, as the quadratic terms lift the accidental blind direction. 
This demonstrates the strong sensitivity to the quadratic terms, which is usually associated with a breakdown of the effective theory expansion and a potential sensitivity to higher-dimensional operators. 
However from \eref{TGC_limits_lep2q} one sees that the new physics scale associated with these operators, e.g. $\Lambda^2\sim m_W^2/g_{1,Z}\approx  (300\textrm{GeV})^2>s_{\textrm{LEP2}}$ , is within the validity of the EFT approach for the energy used at LEP. 
Furthermore even the presence of generic dimension-8 contributions to triple-gauge vertices cannot invalidate the bounds of \eref{TGC_limits_lep2q} \cite{FRinprep} (whether or not this holds when dimension-8 contributions to the t-channel are present, deserves further investigation). 
%The errors are much smaller than for the linear fit in  \eref{TGC_limits_lep2}, as the quadratic terms lift the accidental flat direction. 
%This demonstrates the strong sensitivity to the quadratic terms, which in turn implies a sensitivity to higher-dimensional operators.
%Indeed, in the effective theory approach one implicitly assumes the expansion of observables: 
%$\delta {\cal O} \sim c_{\rm SM}^2 + {s \over v^2} c_{\rm SM} c_{6D}  +  {s^2 \over v^2} c_{6D}^2 + c_{\rm SM}   {s^2 \over v^2}  c_{8D}$, where $c_{6D} \sim v^2/\Lambda^2$ and $c_8 \sim v^4/\Lambda^4 $ represent coefficients of dimension-6 and -8 operators, and $\Lambda$ is the scale of new physics. 
%Thus, $c_{6D}^2$ and $c_{8D}$ terms are formally of the same order in the effective theory expansion in $\Lambda$, and are expected to contribute comparably as long as the SM contribution is not suppressed. 
%In other words,  the results of  the fit in \eref{TGC_limits_lep2q} will be strongly affected if dimension-8 operators are added. 
%
%We conclude that, due to the flat direction,  LEP-2 cannot constrain the parameters of the effective Lagrangian in a completely robust and model independent way.
In any case, and most importantly, \eref{TGC_limits_lep2} contains useful information to constrain concrete new physics models that lead to a subset of dimension-6 operators. 
In particular, in any  model the operator $O_{3W}$ can only be generated at a loop level, therefore the coefficient $c_{3W} = - \lambda_Z$ is suppressed compared to $\delta \hat g_{1,Z}$ and $\delta \hat \kappa_\gamma$ in large classes of models. 
Setting $\lambda_Z = 0$ we obtain
\beq
\label{eq:TGC_limits2_lep2} 
\delta g_{1,Z} = -0.06 \pm 0.03, \quad 
\delta \kappa_{\gamma} = 0.06 \pm 0.04,   \quad  
\rho = \left ( \begin{array}{cc} 
1 & -0.50  \\
\cdot  & 1   
\end{array} \right ) . 
\eeq 
The errors are shrunk by a factor of ten, compared to the general case. 
In this case,  including or not the quadratic terms does not change the result significantly.  
Thus, \eref{TGC_limits_lep2} can be readily used to constrain new physics models predicting $|\lambda_Z| \ll  |\delta \hat g_{1,Z} |, |\delta \hat \kappa_{\gamma}| $; it can be also used when $\lambda_Z$ is not suppressed but is predicted to be away from the blind direction $\lambda_Z \approx - \delta \hat g_{1,Z}$

Finally, we comment on the input from  the LHC. 
One would expect that the LHC may significantly improve on the LEP-2 constraints; in particular, the blind direction, which is plaguing the interpretation of the LEP-2 data, should be lifted. 
So far, ATLAS and CMS have delivered the constraints on the anomalous TGCs  in the WW, WZ, and W$\gamma$ production processes at $\sqrt{s} = 7$~TeV \cite{Chatrchyan:2013yaa, Chatrchyan:2013fya,Aad:2014mda}.  
However, it is difficult to interpret the existing results as constraints on dimension-6 operators in the effective field theory beyond the SM.  
First of all, the experimental collaborations quote the limits only for the case when one or two anomalous TGCs are varied at the same time.
This problem is addressed in Ref.~\cite{Ellis:2014huj}, where a 3-dimensional fit of the anomalous TGCs to the ATLAS 8~TeV  WW distribution is performed. 
However, there is another problem.
The analyses of ATLAS and CMS, as well as that  of  Ref.~\cite{Ellis:2014huj}, focus on the high-$p_T$ tail of the distribution of $W$ and $Z$ decay products, which corresponds to a high center-of-mass energy $\hat s$ of the partonic collision. 
If $\hat s \gtrsim \Lambda$, where $\Lambda \sim v/\sqrt{c_{6D}}$ is the scale suppressing the relevant dimension-6 operators, the process is outside of the range of validity of the EFT.     
We find that this is indeed the case for the magnitude of anomalous TGCs that could produce observable effects in the currently measured LHC distributions. 
Specifically, we find that for $c_{6D} \sim 0.1$, and $c_{8D} \sim c_{6D}^2$,  the contribution of dimension-8 operators to the the events at the high $p_T$ tail exceeds that of dimension-6 operators.  
We conclude that these analyses probe dimension-6 operators in the regime where the EFT expansion is expected to break down; 
in this regard are conclusions are not aligned with those of Ref.~\cite{Ellis:2014huj}. 
However, constraints  derived by these methods may be applied only to concrete models beyond the SM~\cite{Biekoetter:2014jwa}.  
Better designed  observables are needed in order to interpret the $VV$ production at the LHC as model-independent constraints on dimension-6 operators.

%%%%%%%%%%%%%%%%%%%%%%%%%%%%%
\section{Conclusions and Outlook}
\label{sec:conclusions}

This paper discussed in a general way the constraints from  electroweak precision observables  on dimension-6 operators. 
Starting with a redundant set of operators, we identified the combinations that are constrained by the pole observables (W and Z mass and on-shell decays) and by the W boson pair production. 
To this end, we defined a set of effective couplings of W and Z bosons to fermions and to itself, which allow one to consistently include the effects of new physics corrections to gauge boson propagators and vertices. 
These effective couplings are directly related to physical observables, such as the partial decay widths of W and Z bosons or the differential WW production cross section. 
Dimension-6 operators shift the effective couplings away from the SM value, and by calculating this shift one can read off their effect on observables. 
Using this formalism we demonstrated that  the pole observables constrain 8 combinations of dimension-6 operators, while the  W boson pair production constrains another 3 combinations.
We obtained numerical constraints on these combinations in a form that can be easily adapted to any particular basis of operators, or any particular model with new heavy particles.

It is worth stressing that there is a synergy between our precision studies and Higgs precision measurements at the LHC and in future colliders. 
Indeed, most operators in \eref{EFFL_ld6pole} contain the Higgs field, therefore they contribute to Higgs boson decays and/or production. 
Experimental limits on  the coefficient of these operators therefore imply constraints on possible new physics effect in Higgs observables.  
For instance,  along the flat directions of the pole observables there are operators that contribute to the $h\to V \bar f f'$  decays (see e.g. Ref.~\cite{Isidori:2013cla}), and our analysis shows that these are not necessarily tightly constrained.
This example show that constraints from electroweak precision observables may be important in planning the strategy of Higgs measurement.

To derive our results we assumed the coefficients of dimension-6 operators are flavor blind. 
It would be interesting to investigate what happens if this assumption is lifted in a controlled way, for example in the Minimal Flavor Violation scheme\footnote{Ref.~\cite{Pomarol:2013zra} made a first step in this direction and included the leading order in the MFV expansion.}. 
Furthermore, we restricted to observables that do not depend directly on 4-fermion dimension-6 operators.
Lifting this assumption requires dealing with a much larger number of parameters, but also allows one to include many more precision observables, such as fermion scattering off the Z-pole at LEP-2, atomic parity violation, parity-violating electron scattering, etc. 
These directions will be investigated in a future work.

%%%%%%%%%%%%%%%%%%%%%%%%%%%%%%%%%%%%%%
\section*{Acknowledgements}
  
We thank Davide Greco, Martin Jung, Da Liu and Alex Pomarol for useful discussions.
We also thank Alberto Guffanti and Eduard Masso for important comments on the manuscript.
AF~is supported by the ERC Advanced Grant Higgs@LHC. 
FR  acknowledges support from the Swiss National Science Foundation, under the Ambizione grant PZ00P2 136932.

\appendix
%%%%%%%%%%%%%%%%%%%%%%%%%%%%%%%%%%%%%%%%%%%%
\section{Constraints on dimension-6 operators in particular bases}
\label{app:bases}

In the appendix we discuss electroweak precision constraints on the coefficients of dimension-6 operators in three popular bases of operators.   

%---------------------------------------------------------------------
\subsection{Warsaw basis}
\label{app:bases_wb}

In so-called {\em Warsaw basis} of Ref.~\cite{Grzadkowski:2010es}, the set of CP-even operators affecting the pole and WW observables  is chosen as 
\bea
\label{eq:WB_pole}
\cL_{D=6}^{\rm EWPT}  & = & 
 {c_T \over 4 v^2} H^\dagger \overleftrightarrow { D_\mu} H  H^\dagger \overleftrightarrow {D_\mu} H  
 + {c_{WB} \over 4 m_W^2}  B_{\mu\nu}  W_{\mu\nu}^i H^\dagger \sigma^i H  
  + {c_{3W} \over 6 g_L^2 m_W^2} \epsilon^{ijk}    W_{\mu \nu}^{i} W_{\nu\rho}^{j} W_{\rho \mu}^{k} 
\nn  &+ & 
 i {c_{HQ}'  \over v^2} \bar q \sigma^i \bar \sigma_\mu q H^\dagger \sigma^i \overleftrightarrow {D_\mu} H   
+ i{ c_{HQ}  \over v^2}\bar q \bar \sigma_\mu q H^\dagger  \overleftrightarrow {D_\mu} H  
+ i {c_{HU}  \over v^2}  u^c \sigma_\mu \bar u^c  H^\dagger  \overleftrightarrow {D_\mu} H 
+i {c_{HD}  \over v^2}  d^c \sigma_\mu \bar d^c  H^\dagger  \overleftrightarrow {D_\mu} H 
\nn  &+ &  
i {c_{HL}'  \over v^2} \bar \ell \sigma^i \bar \sigma_\mu l H^\dagger \sigma^i \overleftrightarrow {D_\mu} H   
+i {c_{HL} \over v^2} \bar \ell \bar \sigma_\mu l H^\dagger  \overleftrightarrow {D_\mu} H  
+i {c_{HE}  \over v^2} e^c \sigma_\mu \bar e^c  H^\dagger  \overleftrightarrow {D_\mu} H, 
\eea 
Compared to the larger redundant set in \eref{EFFL_ld6pole}, the operators $O_W$, $O_{2W}$, $O_{2B}$, $O_B$ are disposed of via equations of motion \eref{EFFL_eqofmo}, while the operators $O_{HW}$, $O_{HB}$ are removed by integration by parts \eref{EFFL_inbypa}. 
For completeness, we also give the bosonic CP-even operators that only affect Higgs physics, but not the pole observables or gauge boson pair production:  
\bea
\label{eq:WB_higgs}
\cL_{D=6}^{\rm Higgs \ only}  & =  &  
{c_H \over v^2} \, \partial^\mu\!\left( H^\dagger H \right) \partial_\mu \!\left( H^\dagger H \right)   -  c_{6H} \left( H^\dagger H  \right)^3
\nn & + &    \,  
  {c_{GG} \over 4 m_W^2}  H^\dagger H  G_{\mu\nu}^a G^a_{\mu\nu}  
+  { c_{WW} \over 4 m_W^2}   H^\dagger H W_{\mu\nu}^i W_{\mu\nu}^i  
  +  {c_{BB} \over 4 m_W^2}   H^\dagger H B_{\mu\nu}B_{\mu\nu}.
\eea 

Out of the 10 operators in \eref{WB_pole}, the pole observables constrain 7 combinations. 
The constraints can be read off directly from \eref{EWPT_c_pole}: 
\beq
\label{eq:WB_cpole}
\left ( \begin{array}{c} 
c_{HL}'   + c_{WB}  - {g_L^2 \over  4 g_Y^2} c_T   \\
 c_{HL}   -  { 1\over 4} c_T    \\ 
c_{HE}  - {1 \over 2} c_T    \\ 
c_{HQ}'  + c_{WB}  - {g_L^2 \over 4 g_Y^2 } c_T  \\ 
 c_{HQ}  + {1 \over 12} c_T    \\ 
c_{HU} + {1 \over 3} c_T         \\ 
  c_{HD}  - {1 \over 6} c_T            \\   
c_{ll}                           
\end{array} \right )  = 
\left ( \begin{array}{c} 
-1.9 \pm    1.1 \\ 
1.1  \pm   0.7  \\ 
0.1   \pm 0.6    \\ 
-4.7   \pm  1.9   \\ 
 0.2     \pm    2.0  \\ 
7.0          \pm 6.9   \\ 
-31.3          \pm  10.3\\  
- 4.7     \pm  3.5         
\end{array} \right )   \cdot 10^{-3}, \
\eeq 
with the correlation matrix given in \eref{EWPT_c_pole}.  
After applying the pole constraints ,  there are 3 flat directions  among the operators in  \eref{WB_pole} that can be parametrized by $c_{WB}$, $c_T$ and $c_{3W}$. 
From \eref{WB_cpole}, the vertex corrections should be approximately correlated with $c_{WB}$ and $c_T$: 
 \bea & 
\label{eq:WB_flat}
 c_{HL}'  \approx  - c_{WB} +  {g_L^2 c_T \over  4 g_Y^2} , \quad   c_{HL} \approx    {c_T \over 4}, \quad   c_{HE} \approx  {c_T \over 2}, 
& \nn & 
   c_{HQ}'  \approx -  c_{WB} +  {g_L^2 c_T \over 4 g_Y^2 }, \quad  c_{HQ}  \approx  - {c_T \over 12},  \quad 
    c_{HU} \approx - {c_T \over 3}, \quad   c_{HD} \approx   {c_T \over 6} c_T. 
\eea 
These relations should be satisfied at the level of ${\cal O}(10^{-3})$ for the leptonic vertex correction (the first line), and at the level of  ${\cal O}(10^{-2})$ for the quark vertex corrections (the second line). 
The flat directions of the pole observables  are lifted when constraints from gauge boson pair production are taken into account. 
For the sake of studying the constraints from WW production it is a very good approximation to assume that the relations in \eref{WB_flat} hold exactly.  
Then the  relation between the shifts of the effective TGCs in \eref{TGC_dtgc} and the dimension-6 parameters along the pole flat direction is given by 
\beq
\label{eq:WB_tgctod6}
\delta \hat g_{1,Z} =  (g_L^2 +  g_Y^2) \left ( {c_{WB} \over g_L^2} -  {c_T  \over 4 g_Y^2} \right ), 
\quad 
\delta \hat \kappa_{\gamma} = c_{WB},  
\quad 
\lambda_Z  =  -  c_{3W}.  
\eeq 
Rewriting the linear level constraints on anomalous TGCs  in \eref{TGC_limits_lep2} in terms of these dimension-6 operators we obtain
% At the linear level, the LEP-2 WW and single WW production data constrain the dimension-6 operators as 
\beq
\label{eq:WB_lep2tgcfit} 
\left ( \begin{array}{c} 
c_{WB}  \\
c_T   \\ 
c_{3W}             
\end{array} \right )  = 
\left ( \begin{array}{c} 
0.14 \pm    0.05  \\ 
0.86  \pm   0.33  \\ 
-0.86   \pm 0.38       
\end{array} \right ), \quad 
\rho = \left ( \begin{array}{ccc} 
1 & 0.79  & -0.69 \\
\cdot & 1 & -0.99 \\
\cdot & \cdot & 1  
\end{array} \right ) . 
\eeq
In the Warsaw basis the accidental blind direction of LEP-2 occurs along the line $c_T \approx -c_{3W}$. 
The current limits are weak, such that $\cO(1)$ coefficients of dimension-6 operator are allowed by the data. 
In other words, the dimension-6 operators may be suppressed by the scale as small as the weak scale.
This signals a potential sensitivity to dimension-8 and higher operators, if their coefficients take generic value from the EFT point of view.  
However, the constraint are much stronger for away from the blind direction. 
In particular, for $c_{3W} =0$ the constraints following from \eref{WB_lep2tgcfit} reduce to
\beq
\label{eq:WB_lep2tgcfit2} 
\left ( \begin{array}{c} 
c_{WB}  \\
c_T             
\end{array} \right )  = 
\left ( \begin{array}{c} 
0.06 \pm    0.04  \\ 
0.12  \pm   0.06      
\end{array} \right ), \quad 
\rho = \left ( \begin{array}{ccc} 
1 & 0.94 \\ 
\cdot & 1  
\end{array} \right ) . 
\eeq
Note that $c_{WB}$ and $c_T$ can be further  constrained by Higgs data,  together with the operators in \eref{WB_higgs}.

%----------------------------------------------------------------------------------------------------
\subsection{SILH' basis}
\label{app:bases_sp}

The original strongly interacting light Higgs (SILH) Lagrangian \cite{Giudice:2007fh} contains only bosonic operators; 
in  Refs.~\cite{Contino:2013kra} it was extended   to include fermions. 
This precise form is not especially convenient for the sake of  electroweak precision observables because it contains operators $(D_\rho W_{\mu \nu}^a)^2$ and $(\partial_\rho B_{\mu \nu})^2$, which introduce $p^4$ oblique corrections and more complicated tensor structure of the TGCs. 
Here  we use a closely related basis of operators from Refs.~\cite{Elias-Miro:2013mua,Pomarol:2013zra} where these 2 operators are traded for 4-fermion operators. 
We call it the {\em SILH'} basis to distinguish from the original one. 
In the SILH' basis, the operators contributing to the pole observables and to gauge boson pair production are the following: 
\bea
\label{eq:SP_pole}
\cL_{D=6}^{\rm EWPT}  &=&  
 {c_T \over 2} \left (H^\dagger \overleftrightarrow{D_\mu}  H \right) \!\left(   H^\dagger \overleftrightarrow{D_\mu} H\right)  
 + i {v^2 \over m_W^2} \frac{c_W }{2}  H^\dagger  \sigma^a \overleftrightarrow  {D_\mu} H  D_\nu  W_{\mu \nu}^a 
 +i {v^2 \over m_W^2} \frac{c_B}{2} H^\dagger  \overleftrightarrow {D^\mu} H \partial_\nu  B_{\mu \nu}
 \nn 
   &+ &  
i {v^2 \over m_W^2}   c_{HW} D_\mu H^\dagger\sigma^i D_\nu H W^i_{\mu\nu} 
 + i {v^2 \over m_W^2}  c_{HB} D_\mu H^\dagger D_\nu H B_{\mu\nu}
+  {v^2 \over m_W^2} \frac{c_{3W}}{6 g_L^2} \epsilon^{ijk}    W_{\mu \nu}^{i} W_{\nu\rho}^{j} W_{\rho \mu}^{k}
\nn  &+ & 
 i c_{HQ}' \bar q \sigma^i \bar \sigma_\mu q H^\dagger \sigma^i \overleftrightarrow {D_\mu} H   
+i c_{HQ} \bar q \bar \sigma_\mu q H^\dagger  \overleftrightarrow {D_\mu} H  
+ i  c_{HU}  u^c \sigma_\mu \bar u^c  H^\dagger  \overleftrightarrow {D_\mu} H 
+i  c_{HD}  d^c \sigma_\mu \bar d^c  H^\dagger  \overleftrightarrow {D_\mu} H 
\nn  &+ &  
i   c_{HE}  e^c \sigma_\mu \bar e^c  H^\dagger  \overleftrightarrow {D_\mu} H.
\nn
\eea   
For completeness, we also give the CP-even operators that only affect Higgs physics, but not the pole observables or gauge boson pair production:  
\bea
\label{eq:SP_higgs}
\cL_{D=6}^{\rm Higgs \ only}  & =  &  
{c_H \over v^2} \, \partial^\mu\!\left( H^\dagger H \right) \partial_\mu \!\left( H^\dagger H \right)   -  c_{6H} \left( H^\dagger H  \right)^3
\nn & + &    \,  
  {c_{GG} \over 4 m_W^2}  H^\dagger H  G_{\mu\nu}^a G^a_{\mu\nu}  
  +  {c_{BB} \over 4 m_W^2}   H^\dagger H B_{\mu\nu}B_{\mu\nu}.
\eea 
Compared to the Warsaw basis in \eref{WB_pole} and \eref{WB_higgs}, the vertex operators with left-handed leptons $O_{HL}'$ and $O_{HL}$ have been traded for the operators $O_W$ and $O_B$ via the equations of motion  \eref{EFFL_eqofmo}. 
Moreover, the operators $O_{WB}$ and $O_{WW}$ have been traded for $O_{HW}$ and $O_{HB}$ via integration by parts  \eref{EFFL_inbypa}.  

Eight operators in \eref{SP_pole} contribute to the oblique and vertex corrections. 
Seven combinations of those that are constrained by the pole observables can be read off \eref{EWPT_chats}  with $c_{WB} = c_{HL} = c'_{HL}=0$.   
In the SILH' basis the pole observables constrain the parameters as  
\beq
\label{eq:SP_cpole} 
\hspace{-.7cm}
\left ( \begin{array}{c} 
c_T   \\
c_W + c_B    \\ 
c_{HE}    \\ 
c_{HQ}'   \\ 
c_{HQ}    \\ 
c_{HU}          \\ 
c_{HD}             \\   
 c_{ll}                           
\end{array} \right )  = 
\left ( \begin{array}{c} 
-2.2 \pm    1.5 \\ 
- 6.0  \pm   3.4  \\ 
- 2.1   \pm 1.3    \\ 
-2.9   \pm  2.0   \\ 
 0.6     \pm    2.0  \\ 
8.5          \pm 7.0   \\ 
-32.0          \pm  10.4 \\  
 - 4.7     \pm  3.5         
\end{array} \right )   \cdot 10^{-3}, \ 
\rho = \left ( \begin{array}{ccccccccc} 
1 &  0.96 &  0.91 &  -0.34 & -0.12 &   -0.20 &  0.17    & 0.76 \\ 
\cdot   & 1  & 0.95 &  -   0.39 &  -0.13  & - 0.19  &  0.13   & 0.90 \\ 
\cdot   & \cdot &  1 &  -0.46  &  -0.15 &  -0.16  &  0.03  & 0.85  \\ 
\cdot   & \cdot & \cdot & 1 &  -0.29  &  -0.61 &  0.56   & -0.48 \\ 
\cdot   & \cdot & \cdot & \cdot & 1 &   0.44 &  0.19  & -0.14 \\ 
\cdot   & \cdot & \cdot &\cdot &  \cdot &  1 & -0.18  &-0.12   \\ 
\cdot   & \cdot & \cdot & \cdot & \cdot & \cdot & 1  & -0.03 \\
\cdot   & \cdot & \cdot & \cdot & \cdot & \cdot & \cdot  & 1 
\end{array} \right ) . 
\eeq 
Our results somewhat differ from those in Ref.~\cite{Ellis:2014huj} who use this particular basis, which may be due  to a different choice of observables. 
Comparing \eref{WB_cpole} and \eref{SP_cpole} one notes that the constraints on the coefficients of dimension-6 operators are numerically different in the SILH' and in the Warsaw basis.  
The most extreme example is $c_T$, which cannot be constrained by itself in the Warsaw basis, whereas in the SILH' basis it is required to be ${\cal O}(10^{-3})$. 
This is because the same operators can have a different physical interpretation in different bases.  

In the SILH' basis, the pole constraints have a much more intuitive form  than in the Warsaw basis. 
The flat directions of the pole observables can be parameterized by  $c_W$, $c_{HW}$, $c_{HB}$, and $c_{3W}$.  
The remaining parameters in \eref{SP_pole}  are required to be ${\cal O}(10^{-2}-10^{-3})$,  except  for $c_B$ which is constrained by $c_B \approx -c_W$. 
For the sake of studying the constraints from WW production it is a very good approximation to assume these parameter vanish and $c_B = -c_W$ exactly. 
With these assumptions, the shift of the effective TGCs in \eref{TGC_dtgc} in this basis reduce to 
\beq 
 \delta \hat g_{1,Z} = - {g_L^2 + g_Y^2 \over g_L^2} \left (c_W + c_{HW} \right ), 
\quad 
\delta \hat \kappa_\gamma = - c_{HW}  -  c_{HB}
 \quad 
 \lambda_Z =  - \bar  c_{3W} \, .
\label{eq:SILHp_tgc}
\eeq 
Apparently, 4 parameters affect the three TGC shifts that are observable in WW production. 
These constraints can be read off \eref{TGC_limits_lep2} by replacing the anomalous TGCs via   \eref{SILHp_tgc}.
This means that the WW production constraints leave 1 flat direction among  the parameters $c_{HW}$, $c_{HB}$, $c_W$, 
which is an inconvenience of the SILH' basis.
To lift this flat direction one has to include LHC Higgs data, which constrain $c_{HW}$, $c_{HB}$, and $c_W$, as well as  the operators in \eref{SP_higgs}.  

%----------------------------------------------------------------------------------------------------
\subsection{BSM Primaries}
\label{app:bases_BSMP}

As the previous appendices show neither the Warsaw nor the SILH' basis are ideal to compare with experiments: the former due to the large theoretical correlations between $Z$-pole and TGC constraints, the latter due to the correlations between LEP2 and Higgs constraints; furthermore, the very fact they are written in the gauge eigenstate basis (an advantage when comparing with explicit UV models) obscures their impact on physics. 

Refs.~\cite{Gupta:2014rxa,Masso:2014xra} proposes an alternative basis which addresses these problems and is more oriented towards a comparison with experiments: the \emph{BSM Primaries}. It uses field redefinitions to remove all propagator corrections, so that only vertex corrections are left (this makes the implementation in a collider simulator straightforward), and it takes linear combinations of the gauge invariant operators of \eref{EFFL_ld6pole} so that the dimension-6 effects appear in the mass-eigenstate (physical) basis. In this basis the new physics effects corresponding to \eref{EWPT_chats}  can be parametrized through 
 \begin{equation}\label{eq:BSMprimaries}
\{\delta g_{Z\nu}, \delta g_{eZ;L}, \delta g_{eZ;R}, \delta g_{uZ;R}, \delta g_{uZ;L}, \delta g_{dZ;L}, \delta g_{dZ;R}\},
\end{equation}
while modifications to the TGCs can be directly parametrized by $\{ \delta g_1^Z, \kappa_{\gamma}, \lambda_\gamma\}$. The relations of these modifications to other observables (such as Higgs-physics), as implied by the accidental relations of the dimension-6 Lagrangian, can be found in Ref.~\cite{Gupta:2014rxa} and the relation to other bases in Ref.~\cite{Masso:2014xra}. Constraints to the parameters of \eref{BSMprimaries} can be straightforwardly obtained from \eref{EWPT_c_pole} and read\footnote{Notice that Ref.~\cite{Gupta:2014rxa} uses a different normalization (and notation) with $\delta g_{dZ;L}=(\cos\theta_W/g_L)\boldsymbol{\delta g^Z_{dL}}$ etc, which we also adopt here in \eref{BSMP_cpole}.}
\beq
\label{eq:BSMP_cpole} 
\hspace{-.7cm}
\left ( \begin{array}{c} 
\boldsymbol{\delta g^Z_{eL} } \\
\boldsymbol{\delta g^Z_{eR}  } \\ 
\boldsymbol{\delta g^Z_{\nu}   }\\ 
\boldsymbol{\delta g^Z_{uL}  }\\ 
\boldsymbol{\delta g^Z_{dL}  }\\ 
\boldsymbol{\delta g^Z_{uR}   }    \\ 
\boldsymbol{\delta g^Z_{dR}    }         \\   
 c_{ll}                           
\end{array} \right )  = 
\left ( \begin{array}{c} 
0.4 \pm    0.5 \\ 
- 0.1  \pm   0.3  \\ 
- 1.6   \pm 0.8    \\ 
-2.6   \pm  1.6   \\ 
 2.3     \pm    1  \\ 
-3.6          \pm 3.5   \\ 
16.0          \pm  5.2 \\  
 - 4.7     \pm  3.5         
\end{array} \right )   \cdot 10^{-3}, \ 
\rho = \left ( \begin{array}{ccccccccc} 
 1. & 0.66 & -0.43 & -0.16 & 0.16 & 0.02 & -0.14 & -0.43 \\
 \cdot & 1. & -0.02 & -0.01& -0.10& 0.09 & -0.31 & -0.03 \\
 \cdot & \cdot & 1. & 0.03 & -0.02 & 0.01 & 0.00 & 0.96 \\
 \cdot & \cdot & \cdot & 1. & 0.00& 0.70 & -0.25 & -0.01 \\
\cdot & \cdot & \cdot & \cdot & 1. & -0.27 & 0.72 & 0.05 \\
 \cdot & \cdot & \cdot & \cdot & \cdot & 1. & -0.18 & -0.02 \\
 \cdot & \cdot & \cdot & \cdot & \cdot & \cdot & 1. & 0.08 \\
 \cdot & \cdot & \cdot & \cdot & \cdot & \cdot & \cdot & 1. \\
\end{array} \right ) . 
\eeq

\bibliographystyle{JHEP}
\bibliography{TGCplusplus_arxiv3}

\end{document}